\documentclass[runningheads]{llncs}

\usepackage{amsmath,amsfonts}

\usepackage[hidelinks]{hyperref}

\usepackage{algorithmic}
\usepackage{array}

\usepackage{textcomp}
\usepackage{stfloats}
\usepackage{url}
\usepackage{verbatim}
\usepackage{graphicx}
\usepackage{color, colortbl}
\usepackage[table]{xcolor}
\usepackage{enumitem}
\usepackage{amsfonts,amsmath,amssymb,graphicx,setspace}
\usepackage{framed}
\usepackage{multirow}
\usepackage{booktabs}
\usepackage{siunitx}
\usepackage{todonotes}
\usepackage{makecell}
\usepackage{longtable}
\usepackage{xspace}
\usepackage{pgfplots}
\usepackage{pgfplots, pgfplotstable}
\usepackage{bm}
\usepackage{adjustbox}
\usepackage{multicol}
\usepackage{boxedminipage}
\usepackage{multirow}
\usepackage{adjustbox}
\usepackage{multicol}

\def\BibTeX{{\rm B\kern-.05em{\sc i\kern-.025em b}\kern-.08em
    T\kern-.1667em\lower.7ex\hbox{E}\kern-.125emX}}
\usepackage{balance}

\usepackage{pgfplots}

\AtBeginDocument{%
  \providecommand\BibTeX{{%
    Bib\TeX}}}

\usepackage{xspace}

\usepackage{multirow}
\usepackage[mathscr]{eucal}
\usepackage{bm}
\definecolor{Gray}{gray}{0.9}
\usepackage{adjustbox}
\usepgfplotslibrary{groupplots}
\usepgfplotslibrary{colorbrewer}

\usepackage{subcaption}

\definecolor{group1}{HTML}{1F77B4} 
\definecolor{group2}{HTML}{FF7F0E} 
\definecolor{group3}{HTML}{2CA02C} 
\definecolor{group4}{HTML}{D62728} 

\usepgfplotslibrary{colorbrewer}

\usepackage{colortbl}
\usepackage[skins]{tcolorbox}
\newtcolorbox{mybox}[2][]{%
  attach boxed title to top center
               = {yshift=-11pt},
  colframe     =black,
  colbacktitle = black,
  title        = #2,#1,
  enhanced,
}

\newcommand{\et}{\textit{et al.}\xspace}
\usepackage{amsfonts}

\usepackage{enumitem}
\usepackage{lipsum}
\usepackage{xcolor}
 \newcommand{\st}{\scriptscriptstyle}

\usepackage{adjustbox}

\usepackage{multicol}

\usepackage{bm}
\usepackage{multicol}

\newcommand{\cper}{\ensuremath{\bar{\mathtt{\pi}}}\xspace}

\newcommand{\ses}{\ensuremath\mathtt{SS}\xspace}

\newcommand{\se}{\ensuremath{{S}}\xspace}
\newcommand{\re}{\ensuremath{{R}}\xspace}
\newcommand{\p}{\ensuremath{{P}}\xspace}

\newcommand{\tp}{\ensuremath{{T}}\xspace}

\newcommand{\adv}{\ensuremath{\mathcal{A}}\xspace}
\newcommand{\simm}{\ensuremath{\mathtt{Sim}}\xspace}
\newcommand{\view}{\ensuremath{\mathtt{View}}\xspace}
\newcommand{\secsh}{\ensuremath{\mathtt{SS}^{\st(t,n)}}\xspace}
\newcommand{\empt}{\ensuremath{\epsilon}\xspace}

\usepackage{mathtools}
\usepackage{adjustbox}

\usepackage{framed}
\usepackage{tikz}

\newcommand{\ot}{\ensuremath{\mathcal{OT}^{\st 2}_{\st 1}}\xspace}

\usetikzlibrary{arrows,decorations.markings}

\AtBeginDocument{%
  \providecommand\BibTeX{{%
    \normalfont B\kern-0.5em{\scshape i\kern-0.25em b}\kern-0.8em\TeX}}}
    
\begin{document}

\title{Supersonic OT: Fast Unconditionally Secure\\ Oblivious Transfer}


\author{%
Aydin Abadi\thanks{aydin.abadi@ncl.ac.uk}\inst{1} \hspace{1mm} and \hspace{1mm}
Yvo  Desmedt\thanks{y.desmedt@cs.ucl.ac.uk}\inst{2}
\institute{Newcastle University \and The University of Texas at Dallas} 
 }


\maketitle

\begin{abstract}

Oblivious Transfer (OT) is a fundamental cryptographic protocol with applications in secure Multi-Party Computation, Federated Learning, and Private Set Intersection. With the advent of quantum computing, it is crucial to develop unconditionally secure core primitives like OT to ensure their continued security in the post-quantum era. Despite over four decades since OT's introduction, the literature has predominantly relied on computational assumptions, except in cases using unconventional methods like noisy channels or a fully trusted party.

Introducing ``Supersonic OT'', a highly efficient and unconditionally secure OT scheme that avoids public-key-based primitives, we offer an alternative to traditional approaches. Supersonic OT enables a receiver to obtain a response of size $O(1)$. Its simple (yet non-trivial) design facilitates easy security analysis and implementation. The protocol employs a basic secret-sharing scheme, controlled swaps, the one-time pad, and a third-party helper who may be corrupted by a semi-honest adversary. Our implementation and runtime analysis indicate that a single instance of Supersonic OT completes in 0.35 milliseconds, making it up to 2000 times faster than the state-of-the-art base OT.

\end{abstract}


\section{Introduction}\label{sec:intro}


%
%
%

Oblivious Transfer (OT) is a vital cryptographic protocol that enables a user (called a receiver) interested in the $s$-th element of a database  $(m_{\st 0}, m_{\st 1})$  (held by a sender) to learn only $m_{\st s}$ while preserving the privacy of (i) index $s\in\{0, 1\}$ from the sender and (ii) the rest of the database's elements from the receiver. 

OT has found applications within numerous domains, such as generic secure Multi-Party Computation \cite{Yao82b,AsharovL0Z13,HarnikIK07}, Private Set Intersection \cite{DongCW13}, contract signing \cite{EvenGL85}, Federated Learning \cite{YangLCT19,RenYC22,XuLZXND22}, Zero-Knowledge proof systems \cite{GunupudiT08}, and accessing sensitive field elements of remote private databases while preserving privacy \cite{CamenischDN09,AielloIR01,libert2021adaptive}.

As quantum computing emerges, it is crucial to prioritize the development of fundemental security primitives, like OT, with unconditional security. This ensures their sustained resilience in the post-quantum era. 
However, despite over four decades having passed since Rabin introduced OT in 1981 \cite{Rabin-OT}, the security of OTs has predominantly leaned on computational assumptions that might not withstand the power of quantum computers. Efforts have been made to develop \emph{unconditionally secure} OTs, ensuring security even against adversaries armed with quantum computers. Examples include the schemes proposed in \cite{NaorP00,BlundoDSS07,CorniauxG13,CrepeauK88,CrepeauMW04,IshaiKOPSW11,rivest1999unconditionally}. 

However, the state-of-the-art {unconditionally secure} OTs either (i) depend on the multi-sender setting, where each sender possesses a database replica, (ii) utilize a specific communication channel (i.e., noisy channel), or (iii) require the presence of a fully trusted initializer. Nevertheless, distributing the same database across multiple servers, establishing a highly specific communication channel, or involving a fully trusted party would increase the overall deployment cost of these schemes. 

In this work, we propose Supersonic OT, a novel unconditionally secure highly efficient $1$-out-of-$2$ OT that {does not} need to rely on (i) multiple senders, (ii) noisy channel, or (iii) the involvement of a trusted initializer. 
Supersonic OT does not involve any public-key cryptography and allows the receiver to obtain a response of size $O(1)$.  The scheme relies on four main components: a basic secret-sharing scheme \cite{blakley1980one}, a controlled swap \cite{FredkinT02a}, the one-time pad \cite{KatzLindell2014}, and a third-party helper.  This helper could potentially be corrupted by a semi-honest adversary aiming to obtain the private information of the sender or receiver.

The design of Supersonic OT is simple (but elegant), facilitating straightforward security analysis and implementation. Indeed, we have implemented it, made its source code publicly available in \cite{supersonic-code}, and evaluated its overhead. 
A single execution of Supersonic OT completes in approximately 0.35 milliseconds, making it about 1000 times faster than the base OT in \cite{AsharovL0Z13} and up to  $2000$ times faster than the base OT in \cite{Efficient-OT-Naor}.

\section{Preliminaries}

\subsection{Notations}\label{sec::notations}
By $\empt$ we mean an empty string. We denote a sender by $\se$ and a receiver by $\re$. We assume parties interact with each other through a regular secure channel. By the notation, $\mathcal{X} \stackrel{c}{\equiv} \mathcal{Y}$, we mean that the two distributions $\mathcal{X}$ and $\mathcal{Y}$ are computationally indistinguishable. By  $\mathcal{X} {\equiv}  \mathcal{Y}$ we mean $\mathcal{X}$ and $\mathcal{Y}$ are unconditionally indistinguishable.  
$U$ denotes a universe of messages $m_{\st 1},\ldots, m_{\st t}$. We define $\sigma$ as the maximum size of messages in $U$, i.e., $\sigma=Max(|m_{\st 1}|,\ldots, |m_{\st t}|)$. 
%


\subsection{Security Model}\label{sec::sec-model}

In this paper, we rely on the simulation-based model of secure multi-party computation \cite{DBLP:books/cu/Goldreich2004} to define and prove the proposed protocols. Below, we restate the formal security definition within this model.

 \subsubsection{Two-party Computation.} A two-party protocol $\Gamma$ problem is captured by specifying a random process that maps pairs of inputs to pairs of outputs, one for each party. Such process is referred to as a functionality denoted by  $f:\{0,1\}^{\st  *}\times\{0,1\}^{\st  *}\rightarrow\{0,1\}^{ \st *}\times\{0,1\}^{ \st *}$, where $f:=(f_{\st  1},f_{\st  2})$. For every input pair $(x,y)$, the output pair is a random variable $(f_{\st  1} (x,y), f_{\st  2} (x,y))$, such that the party with input $x$ wishes to obtain $f_{\st  1} (x,y)$ while the party with input $y$ wishes to receive $f_{\st  2} (x,y)$. 
%
 In the setting where $f$ is asymmetric and only one party (say the first one) receives the result, $f$ is defined as $f:=(f_{\st  1}(x,y), \empt)$.

 \subsubsection{Security in the Presence of Passive Adversaries.}  In the passive adversarial model, the party corrupted by such an adversary correctly follows the protocol specification. Nonetheless, the adversary obtains the internal state of the corrupted party, including the transcript of all the messages received, and tries to use this to learn information that should remain private. Loosely speaking, a protocol is secure if whatever can be computed by a party in the protocol can be computed using its input and output only. In the simulation-based model, it is required that a party’s view in a protocol's 
 execution can be simulated given only its input and output. 
 
 This implies that the parties learn nothing from the protocol's execution. More formally, party $i$’s view (during the execution of $\Gamma$) on input pair  $(x, y)$ is denoted by $\mathsf{View}_{\st  i}^{\st  \Gamma}(x,y)$ and equals $(w, r_{\st  i}, m_{\st  1}^{\st  i}, \ldots, m_{\st  t}^{\st  i})$, where $w\in\{x,y\}$ is the input of $i^{\st  th}$ party, $r_{\st  i}$ is the outcome of this party's internal random coin tosses, and $m_{\st  j}^{\st  i}$ represents the $j^{\st  th}$ message this party receives.  The output of the $i^{\st  th}$ party during the execution of $\Gamma$ on $(x, y)$ is denoted by $\mathsf{Output}_{\st  i}^{\st  \Gamma}(x,y)$ and can be generated from its own view of the execution.  
\vspace{-1mm}
\begin{definition}\label{def::sim-def}
Let $f$ be the deterministic functionality defined above. Protocol $\Gamma$ securely computes $f$ in the presence of a  passive adversary if there exist polynomial-time algorithms $(\mathsf {Sim}_{\st  1}, \mathsf {Sim}_{\st  2})$ such that:
  \begin{equation*}
  \{\mathsf {Sim}_{\st 1}(x,f_{\st 1}(x,y))\}_{\st x,y}{\equiv} \{\mathsf{View}_{\st 1}^{\st \Gamma}(x,y) \}_{\st x,y}
  \end{equation*}
  \begin{equation*}
    \{\mathsf{Sim}_{\st 2}(y,f_{\st 2}(x,y))\}_{ \st x,y} {\equiv} \{\mathsf{View}_{\st 2}^{\st \Gamma}(x,y) \}_{\st x,y}
  \end{equation*}
 \end{definition}

Definition \ref{def::sim-def} supports  
unconditional indistinguishably. In order to support computational indistinguishably, we can substitute ${\equiv}$ with $\stackrel{c}{\equiv}$. 



\subsection{Controlled Swap}\label{Customised-Random-Swap}
The idea of the controlled swap was introduced by Fredkin and Toffoli  \cite{FredkinT02a}. It can be defined as function $\cper(.)$ which takes two inputs: a binary value $s$ and a pair $(c_{\st 0}, c_{\st 1})$. When $s=0$, it returns the input pair $(c_{\st 0}, c_{\st 1})$, i.e., it does not swap the elements. When $s=1$, it returns $(c_{\st 1}, c_{\st 0})$, effectively swapping the elements. 
It is evident that if $s$ is uniformly chosen at random, then $\cper(.)$ represents a random permutation, implying that the probability of swapping or not swapping is $\frac{1}{2}$. We will use $\bar{\pi}(.)$ in the protocol presented in Figure \ref{fig::Ultrasonic-OT}.

\subsection{Secret Sharing}\label{sec::secret-haring}

A (threshold) secret sharing $\mathtt{SS}^{\st(t,n)}$  scheme is a cryptographic protocol that enables a dealer to distribute a string $s$, known as the secret, among $n$ parties in a way that the secret $s$ can be recovered when at least a predefined number of shares, say $t$, are combined. If the number of shares in any subset is less than $t$, the secret remains unrecoverable and the shares divulge no information about $s$. This type of scheme is referred to as $(n, t)$-secret sharing or \secsh for brevity. 

In the case where $t=n$, there exists a highly efficient XOR-based secret sharing \cite{blakley1980one}. In this case, to share the secret $s$, the dealer first picks $n-1$ random bit strings $r_{\st 1}, \ldots, r_{\st n-1}$ of the same length as the secret. Then, it computes $r_{\st n} = r_{\st 1} \oplus \ldots \oplus  r_{\st n} \oplus s$. It considers each $r_{\st i}\in\{r_{\st 1}, \ldots,r_{\st n}\}$ as a share of the secret. To reconstruct the secret, one can easily compute $r_{\st 1}\oplus\ldots \oplus\ r_{\st n}$. Any subset of less than $n$ shares reveals no information about the secret. We will use this scheme in this paper. A secret sharing scheme involves two main algorithms; namely, $\ses(1^{\st \lambda}, s, n, t)\rightarrow (r_{\st 1}, \ldots, r_{\st n})$: to share a secret and $\mathtt{RE}(r_{\st 1}, \ldots, r_{\st t}, n, t)\rightarrow s$ to reconstruct the secret.


\section{Related Work}\label{sec::related-work}


 Oblivious Transfer (OT) is one of the important building blocks of cryptographic protocols and has been used in various mechanisms. 
 The traditional $1$-out-of-$2$ OT (\ot) is a protocol that involves two parties, a sender \se and a receiver \re.  \se has a pair of input messages $(m_{\st 0}, m_{\st 1})$ and \re has an index $s$. The aim of \ot is to enable \re to obtain $m_{\st s}$, without revealing anything about $s$ to \se, and without allowing \re to learn anything about  $m_{\st 1-s}$. The traditional \ot functionality is defined as $\mathcal{F}_{\st\ot}:((m_{\st 0}, m_{\st 1}), s) \rightarrow (\empt, m_{\st s})$.   
  
The notion of $1$-out-of-$2$ OT was initially proposed by Rabin \cite{Rabin-OT} which consequently was generalized by Even \et \cite{EvenGL85}. Since then, numerous variants of OT have been proposed. For instance, 

\begin{itemize}[label=$\bullet$]
\item $1$-out-of-$n$ OT, e.g., in \cite{NaorP99,Tzeng02,LiuH19}: which allows \re to pick one entry out of $n$ entries held by \se, 

\item $k$-out-of-$n$ OT, e.g., in \cite{ChuT05,JareckiL09,ChenCH10}: which allows \re to select $k$ entries out of $n$ entries held by \se,

\item OT extension, e.g., in \cite{IshaiKNP03,Henecka013,Nielsen07,AsharovL0Z13}: that supports efficient executions of OT (that mainly relies on symmetric-key operations), in the case OT needs to be invoked many times
 
\item distributed OT, e.g., in \cite{NaorP00,CorniauxG13,ZhaoSJMZX20}: that allows the database to be distributed among $m$ servers/senders. 

\end{itemize}

In the remainder of this section, we discuss those variants of OT  that are closer to our work. We refer readers to \cite{OT-Survey} for a recent survey of OT.

\subsection{Unconditionally and Post-Quantum Secure OTs}\label{sec::uncon-sec-OT} There have been efforts to design (both-sided) unconditionally secure OTs. Some schemes, such as those proposed in  \cite{NaorP00,BlundoDSS07,CorniauxG13}, rely on multiple servers/senders that maintain an identical copy of the database.  Other ones, like the one introduced in \cite{CrepeauK88,CrepeauMW04,IshaiKOPSW11}, are based on a specific network structure, i.e., a noisy channel, to achieve unconditionally secure OT.   There is also a scheme in \cite{rivest1999unconditionally} that achieves unconditionally secure OT using a fully trusted initializer.

Moreover, there exist OT schemes developed to maintain security in the presence of adversaries equipped with quantum computers. Examples include those proposed in \cite{BlazyCV19,BlazyC15,PeikertVW08,kundu20201,DowsleyGMN12,BarretoOB18}. 
However, these schemes are not unconditionally secure. Instead, they rely on various assumptions and problems (such as short integer solution, learning with errors, multivariate quadratic, decoding random linear codes, or computing isogenies between supersingular elliptic curves) as well as primitives (such as AES, pseudorandom generator, lattice-based Chameleon hash function, multivariate quadratic cryptography, McEliece cryptosystem, or supersingular isogeny Diffie-Hellman key exchange) that are deemed valid and secure in the era of quantum computing based on current knowledge and assessment. Their security could be compromised if any of the underlying assumptions or problems are proven to be solvable efficiently by future advancements in quantum algorithms or other unforeseen developments.

Hence, there exists no (efficient) unconditionally secure OT that does not use noisy channels, multi-server, and fully trusted initializer. 

\subsection{OT with Constant Response Size}

Researchers have proposed several OTs, e.g., those proposed in \cite{CamenischNS07,GreenH08,ZhangLWR13},  that enable a receiver to obtain a constant-size response to its query. To achieve this level of communication efficiency, these protocols require the receiver to locally store the encryption of the entire database, in the initialization phase. During the transfer phase, the sender assists the receiver with locally decrypting the message that the receiver is interested in. 

The main limitation of these protocols is that a thin client with limited available storage space cannot use them, as it cannot locally store the encryption of the entire database.

\section{Supersonic OT}\label{sec::supersonice-OT}



In this section, we introduce a $1$-out-of-$2$ OT, called ``Supersonic OT'',  which (i)
operates at high speed by eliminating the need for public-key-based cryptography, (ii) delivers a response of size $O(1)$ to the recipient, \re, and (iii) ensures information-theoretic security, making it post-quantum secure.

\subsection{Security Definition}\label{sec::Ultra-OT-definition}

Supersonic OT involves three types of entities, a sender \se, a receiver \re, and a helper \p. We assume each party can be corrupted by a passive non-colluding adversary. The functionality $\mathcal{F}_{\scriptscriptstyle\ot}$ that Supersonic OT will compute is similar to that of conventional OT with the difference that now an additional party $\p$ is introduced, having no input and receiving no output. Thus, we define the functionality as   $\mathcal{F}_{\scriptscriptstyle\ot}:\big((m_{\st 0}, m_{\st 1}), \empt, s\big) \rightarrow (\empt, \empt, m_{\st s})$. 
 Next, we present a formal definition of \ot.

\begin{definition}[\ot]\label{def::ultra-OT-sec-def} Let $\mathcal{F}_{\scriptscriptstyle\ot}$ be the   OT functionality defined above. We assert that protocol $\Gamma$ realizes $\mathcal{F}_{\scriptscriptstyle\ot}$ in the presence of a passive adversary, if for  every adversary \adv
in the real model, there is a simulator \simm  in
the ideal model, where:

\begin{equation}\label{equ::ultra-ot-sender-sim-}
\begin{split}
\Big\{\simm_{\st\se}\big((m_{\st 0}, m_{\st 1}), \empt\big)\Big\}_{\st m_{0}, m_{1}, s}{\equiv}\Big\{\view_{\st\se}^{ \st\Gamma}\big((m_{\st 0}, m_{\st 1}), \empt,  s\big) \Big\}_{\st m_0, m_1, s}
\end{split}
\end{equation}

\begin{equation}\label{equ::ultra-ot-server-sim-}
\begin{split}
\Big\{\simm_{\st\p}(\empt, \empt)\Big\}_{\st m_0, m_1, s}{\equiv}  \Big\{\view_{\st\p}^{\st \Gamma}\big((m_{\st 0}, m_{\st 1}), \empt, s\big) \Big\}_{\st m_0, m_1, s}
\end{split}
\end{equation}

\begin{equation}\label{equ::ultra-ot-reciever-sim-}
\begin{split}
\Big\{\simm_{\st\re}\Big(s, \mathcal{F}_{\scriptscriptstyle\ot}\big((m_{\st 0}, m_{\st 1}), \empt,  s\big)\Big)\Big\}_{\st m_0, m_1, s}{\equiv}  \Big\{\view_{\st\re}^{\st \Gamma}\big((m_{\st 0}, m_{\st 1}), \empt,   s\big) \Big\}_{\st m_0, m_1, s}
\end{split}
\end{equation}

\end{definition}

\subsection{The Protocol}

At a high level, the protocol operates as follows. Initially, \re and \se agree on a pair of keys. In the query generation phase, \re splits its private index into two binary shares. It sends one share to \se and the other to \p. Given the share/query, \se encrypts every message $m_{\st i}$ (using a one-time pad) under one of the keys it agreed with \re.  

Accordingly, \se permutes the encrypted messages using  $\cper$ and its share. It sends the resulting pair to \p which permutes the received pair using  $\cper$ and its share. \p sends only the first element of the resulting pair (which is a ciphertext) to \re and discards the second element of the pair. Consequently, \re decrypts the ciphertext and learns the message it was interested in. Figure \ref{fig::Ultrasonic-OT} presents Supersonic OT in detail. 

As it is evident, \p plays a minimal role, involving permuting and obliviously filtering out the message it receives from \se.  The size of the single message that \re receives can be short (e.g., 64 or 128 bits) depending on the maximum bit size of the messages that \se holds and the security parameter.

%
\begin{figure}[!h]
\begin{center}
    \begin{tcolorbox}[enhanced,  
    drop fuzzy shadow southwest,
    colframe=black,colback=white]
\begin{enumerate}

\item \underline{\textit{\re-side Setup:}}
$\mathtt{Setup}(1^{\st\lambda})\rightarrow (k_{\st 0}, k_{\st 1})$ 

\begin{itemize}


\item[$\bullet$]  picks two random keys $(k_{\st 0}, k_{\st 1}) \stackrel{\st\$}\leftarrow\{0, 1\}^{\st\sigma}$ and sends them to \se.

\end{itemize}

\item \underline{\textit{$\re$-side Query Generation:}} 
$\mathtt{GenQuery}(1^{\st \lambda}, s)\rightarrow q=(q_{\st 1}, q_{\st 2})$

\begin{enumerate}

\item splits  the private index $s$ into two shares $(s_{\st 1}, s_{\st 2})$ by calling:
  $$\ses(1^{\st \lambda}, s, 2, 2)\rightarrow (s_{\st 1}, s_{\st 2})$$ 

\item sends $q_{\st 1}=s_{\st 1}$ to \se and $q_{\st 2}=s_{\st 2}$ to $\p$. 

\end{enumerate}

\item \underline{\textit{$\se$-side Response Generation:}} 
$\mathtt{GenRes}(m_{\st 0}, m_{\st 1}, k_{\st 1},k_{\st 2}, q_{\st 1})\rightarrow res$

\begin{enumerate}

\item encrypts each message as follows. 
%
$$\forall i, 0\leq i\leq 1: m'_{\st i}=m_{\st i}\oplus k_{\st i}$$

Let $e=(m'_{\st 0}, m'_{\st 1})$ contain the encrypted messages. 

\item permutes the elements of $e$: 
$$\cper(s_{\st 1}, e)\rightarrow e'$$

\item sends $res=e'$ to $\p$. 

\end{enumerate}

\item \underline{\textit{$\p$-side Oblivious Filtering:}} 
$\mathtt{OblFilter}(res, q_{\st 2})\rightarrow res'$

\begin{enumerate}

\item permutes the elements of $e'$: 
$$\cper(s_{\st 2}, e')\rightarrow e''$$

\item\label{ultra-ot::e-double-prime} sends (always) the first element in $e''$, say $res'=e''_{\st 0}$, to \re and discards the second element in $e''$. 

\end{enumerate}

\item\underline{\textit{\re-side Message Extraction:}} 
$\mathtt{Retreive}(res',  k_{\st s})\rightarrow m_{\st s}$

\begin{itemize}
\item[$\bullet$] retrieves the final related message $m_{\st s}$ by decrypting $e''_{\st 0}$: 
$$m_{\st s}=e''_{\st 0}\oplus k_{\st s}$$
\end{itemize}

\end{enumerate}
\end{tcolorbox}
\end{center}
\vspace{-2mm}
\caption{Supersonic OT.} 
\vspace{-2mm}
\label{fig::Ultrasonic-OT}
\end{figure}



\begin{theorem}\label{theo::ultra-OT}
Let $\mathcal{F}_{\scriptscriptstyle\ot}$ be the functionality defined in Section \ref{sec::Ultra-OT-definition}.  Then, Supersonic OT (presented in Figure \ref{fig::Ultrasonic-OT}) securely computes $\mathcal{F}_{\scriptscriptstyle\ot}$ in the presence of semi-honest adversaries, 
%
%
w.r.t. Definition \ref{def::ultra-OT-sec-def}. 
\end{theorem}

In the following subsections, we will prove the above theorem and then prove the correctness of Supersonic OT. 



\vspace{8mm}
\subsection{Proof of Security}\label{sec::ultrasonic-ot-proof}

In this section, we prove the security of Supersonic OT, i.e., Theorem \ref{theo::ultra-OT}. 

\begin{proof}
We consider the case where each party is corrupt at a time.

\subsubsection{Corrupt Receiver \re.} In the real execution, \re's view is defined as: 
$$\view_{\st\re}^{\st Supersonic\text{-}OT}\big((m_{\st 0}, m_{\st 1}), \empt,  s\big) = \{r_{\st\re}, e''_{\st 0}, m_{\st s} \}$$

where $r_{\st \re}$ is the outcome of the internal random coin of  \re and is used to generate $(s_{\st 1}, s_{\st 2}, k_{\st 0}, k_{\st 1})$.  
 Below, we construct an idea-model simulator $\simm_{\st \re}$ which receives $(s, m_{\st s})$ from \re.

\begin{enumerate}
\item initiates an empty view and appends a uniformly random coin $r'_{\st \re}$ to the view. 

\item picks a random key $k \stackrel{\st\$}\leftarrow\{0, 1\}^{\st\sigma}$, using $r'_{\st \re}$. 

\item encrypts  message $m_s$ as follows $e=m_s\oplus k$. 

\item appends $e$ to the view and outputs the view. 

\end{enumerate}

Since we are in the passive adversarial model, the adversary picks its random coin $r_{\st \re}$  (in the real models) according to the protocol. Therefore, $r_{\st \re}$ and $r'_{\st \re}$ have identical distributions. Moreover, $e''_{\st 0}$ in the real model and $e$ in the ideal model have identical distributions as both are the result of XORing message $m_s$ with a fresh uniformly random value. Also, $m_s$ is the same in both models so it has identical distribution in the real and ideal models. We conclude that Relation \ref{equ::ultra-ot-reciever-sim-} (in Section \ref{sec::Ultra-OT-definition}) holds.

\subsubsection{Corrupt Sender \se.}
In the real execution, \se's view can be defined as: 
$$\view_{\se}^{\st Supersonic\text{-}OT}\big((m_{\st 0}, m_{\st 1}), \empt,   s\big) = \{r_{\st\se}, s_{\st 1}, k_{\st 0}, k_{\st 1} \}$$

 where $r_{\st \se}$ is the outcome of the internal random coin of  \se and is used to generate its random values.   Next, we construct an idea-model simulator $\simm_{\st\se}$ which receives $(m_{\st 0}, m_{\st 1})$ from \se.

\begin{enumerate}
\item initiates an empty view and appends a uniformly random coin $r'_{\st \se}$ to the view. 

\item picks a binary random value $s' \stackrel{\st\$}\leftarrow\{0, 1\}$.

\item picks two uniformly random keys $(k'_{\st 0}, k'_{\st 1}) \stackrel{\st\$}\leftarrow\{0, 1\}^{\st\sigma}$.

\item  appends $s', k'_{\st 0}, k'_{\st 1}$ to the view and outputs the view. 
\end{enumerate}
Next, we explain why the two views are indistinguishable. 
The random coins $r_{\st \se}$ and $r'_{\st \se}$ in the real and ideal models have identical distributions as they have been picked according to the protocol's description (as we consider the passive adversarial model). Moreover, $s_{\st 1}$ in the real model and $s'$ in the ideal model are indistinguishable, as due to the security of the secret sharing scheme, binary share $s_{\st 1}$ is indistinguishable from a random binary value $s'$. Also, the elements of pair $(k_{\st 0}, k_{\st 1})$ in the real model and the elements of pair $(k'_{\st 0}, k'_{\st 1})$ in the ideal model have identical distributions as they have been picked uniformly at random from the same domain. Hence,  Relation \ref{equ::ultra-ot-sender-sim-} (in Section \ref{sec::Ultra-OT-definition}) holds.

\subsubsection{Corrupt Helper \p.}
In the real execution, \p's view is defined as: 
$$\view_{\st\p}^{\st Supersonic\text{-}OT}\big((m_{\st 0}, m_{\st 1}), \empt,   s\big) = \{r_{\st\p}, s_{\st 2}, e' \}$$

   where $r_{\st \p}$ is the outcome of the internal random coin of  \p and is used to generate its random values and  $e'$ is a pair $(e'_{\st 0}, e'_{\st 1})$ and is an output of $\cper$.    Next, we construct an idea-model simulator $\simm_{\st\p}$.

\begin{enumerate}
\item initiates an empty view and appends a uniformly random coin $r'_{\st \p}$ to the view. 

\item picks a binary random value $s' \stackrel{\st\$}\leftarrow\{0, 1\}$.

\item constructs a pair $v$ of  two uniformly random values $(v_{\st0}, v_{\st1}) \stackrel{\st\$}\leftarrow\{0, 1\}^{\st\sigma}$.

\item appends $s', v$ to the view and outputs the view. 

\end{enumerate}

Since we consider the passive adversarial model, the adversary picks its random coins $r_{\st \p}$ and $r'_{\st \p}$ (in the real and ideal models respectively) according to the protocol. So, they have identical distributions. Moreover, $s_{\st 2}$ in the real model and $s'$ in the ideal model are indistinguishable, as due to the security of the secret sharing scheme, binary share $s_{\st 2}$ is indistinguishable from a random binary value $s'$. 
In the real model, the elements of $e'$  which are $e'_{\st 0}$ and $e'_{\st 1}$ have been encrypted/padded with two fresh uniformly random values. In the ideal model, the elements of $v$ which are $v_{\st0}$ and $v_{\st 1}$ have been picked uniformly at random. Due to the security of a one-time pad, $e'_{\st i}$ ($\forall i, 0\leq i\leq 1$) is indistinguishable from a uniformly random value, including $v_{\st 0}$ and $v_{\st 1}$.  

Also, in the real model, the pair $e'$ that is given to \p  is always permuted based on the value of \se's share (i.e., $s_{\st 1}\in\{0, 1\}$) which is not known to \p; whereas, in the ideal model, the pair $v$ is not permuted. However, given the permuted pair $e'$ and not permuted pair $v$, a distinguisher cannot tell where each pair has been permuted with a probability greater than $\frac{1}{2}$.   Therefore,  Relation \ref{equ::ultra-ot-server-sim-} (in Section \ref{sec::Ultra-OT-definition}) holds. 
\hfill\(\Box\) 
 \end{proof}


\subsection{Proof of Correctness}\label{sec::Ultra-OT-Proof-of-Correctness}


In this section, we demonstrate that \re always receives the message $m_{\st s}$ corresponding to its query $s$. To accomplish this, we will show that (in step \ref{ultra-ot::e-double-prime}) the first element of pair $e''$ always equals the encryption of $m_s$. This outcome is guaranteed by the following two facts: (a) $s=s_{\st 1}\oplus s_{\st 2}$ and (b)  \se and \tp permute their pairs based on the value of their share, i.e., $s_{\st 1}$ and $s_{\st 2}$ respectively. 


\begin{table}[!htb]
\vspace{-2mm}
\begin{center}
\scalebox{1}{
  \begin{tabular}{|c|c|c|c|c|}   
\hline
    $s$&$s_{\st 1}$&$s_{\st 2}$\\   
\hline

\hline
&\cellcolor{gray!20}1&\cellcolor{gray!20} 1\\

\cline{2-3}

\multirow{-2}{*}{$0$}&\cellcolor{gray!20}0&\cellcolor{gray!20}0\\

\hline

&\cellcolor{gray!20}1&\cellcolor{gray!20}0\\

\cline{2-3}

\multirow{-2}{*}{$1$}&\cellcolor{gray!20}0&\cellcolor{gray!20}1\\

\hline

\end{tabular}
}
\end{center}
\caption{Relation between query $s$ and behaviour of permutation $\cper$ from the perspective of $\se$ and $\p$. When $s_{\st i}=1$, $\cper$ swaps the elements of its input pairs and when $s_{\st i}=0$, $\cper$ does not swap the elements of the input pairs.}
\label{tab:ultra-ot-correctness}
\vspace{-3mm}
\end{table}


As Table \ref{tab:ultra-ot-correctness} indicates, when $s=0$, then (i) either both $\se$ and \p permute their pairs or (ii) neither does. In the former case, since both swap the elements of their pair, then the final permuted pair $e''$ will have the same order as the original pair $e$ (before it was permuted). In the latter case, again $e''$ will have the same order as the original pair $e$ because neither party has permuted it. Thus, in both of the above cases (when $s=0$), the first element of $e''$ will be the encryption of $m_{\st 0}$. 
Moreover, as Table \ref{tab:ultra-ot-correctness} indicates, when $s=1$, then only one of the parties \se and \p will permute their input pair. This means that the first element of the final permuted pair $e''$  will always equal the encryption of $m_{\st 1}$. 
%

\section{Performance Evaluation}\label{sec::supersonic-OT}

We have implemented Supersonic OT in C++ and evaluated its concrete runtime. The source code for the implementation is publicly available in \cite{supersonic-code}. For the experiment, we utilized a MacBook Pro laptop equipped with a quad-core Intel Core i5, 2 GHz CPU, and 16 GB RAM. Given the low number and small size of exchanged messages in our protocol, implementing it in a network setting will yield negligible network overhead.  
We did not leverage parallelization or any other optimization. The experiment was run an average of 50 times. We utilized the GMP library \cite{gmp} for big-integer arithmetic.

We analyzed the runtime of various phases of Supersonic OT across different invocation frequencies ($1, 10, 10^{\st 3}, 10^{\st 5}$, and $10^{\st 7}$ times). Table \ref{fig::supersonic-cost-breakdown-by-phases} shows the high-speed performance of Supersonic OT, requiring only $0.35$ milliseconds (ms) for a single invocation. Notably, Phase 1 incurs the highest computation cost when the number of invocations is $1, 10$, and $10^{\st 3}$, while Phase 3 imposes the highest computation cost when the number of invocations is $10^{\st5}$ and $10^{\st7}$. Overall, Phase 2 has the lowest computation cost.


\begin{table}[!htbp]
\begin{center}
\vspace{-3mm}
\caption{ The table presents the runtime of Supersonic OT, categorized by various phases and measured in ms.  The security parameter and message size are set at 128 bits.}  \label{fig::supersonic-cost-breakdown-by-phases}

\renewcommand{\arraystretch}{1.5}

\begin{tabular}{|c|c|c|c|c|c|c|c|c|c|c|c|} 
\cline{1-7}

\multirow{2}{*} {  Protocol}&\multirow{2}{*} {  Phases}&

 \multicolumn{5}{c|}{  Number of OT Invocations}\\

            \cline{3-7}
  &&\cellcolor{white!20} $1$&\cellcolor{white!20}  $10$&\cellcolor{white!20} $10^{\st 3}$&\cellcolor{white!20} $10^{\st 5}$&\cellcolor{white!20} $10^{\st 7}$\\
            
    \cline{1-2}

    \hline

    \multirow{6}{*}{\rotatebox[origin=c]{90}{    Supersonic OT }}&\cellcolor{white!20}  Phase 1&\cellcolor{gray!50}  0.34   &\cellcolor{gray!50}   0.37&\cellcolor{gray!50}    0.61&\cellcolor{gray!50}   21.17 &\cellcolor{gray!50}   1990\\
    
       \cline{2-7}

&\cellcolor{white!20}  Phase 2&\cellcolor{gray!20}   0.00069&\cellcolor{gray!20}   0.00092&\cellcolor{gray!20}   0.0071 &\cellcolor{gray!20}   0.7& \cellcolor{gray!20}   63.86 \\ 

   \cline{2-7}
     
&\cellcolor{white!20}  Phase 3&\cellcolor{gray!50}  0.0011 &\cellcolor{gray!50}   0.0038&\cellcolor{gray!50}   0.29 &\cellcolor{gray!50}  29.48  & \cellcolor{gray!50}  3058.09\\ 

       \cline{2-7}
          
&\cellcolor{white!20}  Phase 4&\cellcolor{gray!20}   0.00065&\cellcolor{gray!20}   0.0011&\cellcolor{gray!20}    0.061&\cellcolor{gray!20}   6.3& \cellcolor{gray!20}  827.59\\ 

       \cline{2-7}

&\cellcolor{white!20}  Phase 5&\cellcolor{gray!50}   0.00064&\cellcolor{gray!50}   0.0012&\cellcolor{gray!50}   0.063 &\cellcolor{gray!50}  6.016 & \cellcolor{gray!50}   675.06 \\ 

       \cline{2-7}
        \cline{2-7}

&\cellcolor{white!20}  Total&\cellcolor{gray!20}   0.35&\cellcolor{gray!20}   0.38&\cellcolor{gray!20}   1.05 &\cellcolor{gray!20}  63.7& \cellcolor{gray!20}   6610 \\ 

    \hline

\end{tabular}
\end{center}
\vspace{-2mm}
\end{table}

\vspace{-3mm}
\subsection{Runtime}



\subsubsection{Supersonic Versus Base OTs.}

Initially, we compare the runtime of Supersonic OT with that of \textit{base} OTs proposed in \cite{AsharovL0Z13,Efficient-OT-Naor}. These base OTs, known for their generality, efficiency, and widespread usage in literature, serve as efficient foundations in OT extensions. 
Table \ref{fig::supersonic-vs-base-OTs} and Figure \ref{supersonic-vs-base} summarize this comparison. The runtime data for OTs in \cite{AsharovL0Z13,Efficient-OT-Naor} is derived from the figures reported in \cite{AsharovL0Z13}, specifically from Table 3, where the GMP library was employed. \footnote{The implementation in \cite{AsharovL0Z13} used a smaller RAM size (4 GB) compared to our usage of 16 GB. Their reported performance also accounts for LAN delay. It is important to note that doubling the RAM size doesn't necessarily equate to doubling the runtime. Moreover, the impact of LAN delay is negligible, particularly for our OT, which incurs a very low communication cost.}
Table \ref{fig::supersonic-vs-base-OTs} highlights that Supersonic OT demonstrates a speed advantage, being approximately $10^{\st 3}$ times faster than the OT in \cite{AsharovL0Z13} and up to around $2\times 10^{\st 3}$ times faster than the OT in \cite{Efficient-OT-Naor}.


\begin{table}[!htbp]
\vspace{-3mm}
\caption{ The table compares the runtime of Supersonic OT with the following base OTs: standard OT (STD--OT) in  \cite{AsharovL0Z13}, STD--OT in \cite{Efficient-OT-Naor}, and the random oracle OT (RO--OT) in \cite{Efficient-OT-Naor}.  The bit size of the security parameter is 128. The runtime has been measured in ms and is based on 128 invocations of each scheme. The enhancement ratio refers to the performance improvement that Supersonic OT offers in comparison to each specific scheme.}  
\label{fig::supersonic-vs-base-OTs}
\begin{center}
\renewcommand{\arraystretch}{1.5}

\begin{tabular}{|c|c|c|c|c|c|c|c|c|c|c|c|} 
    \cline{1-3}

    Scheme&    Runtime&    Enhancement Ratio\\

       \hline

\cellcolor{white!20}   STD--OT in \cite{AsharovL0Z13}&\cellcolor{gray!50}   1,217  &\cellcolor{gray!50}     1,622\\
    
        \hline

\cellcolor{white!20}   STD--OT in \cite{Efficient-OT-Naor}&\cellcolor{gray!20}    1,681&\cellcolor{gray!20}     2,241\\ 

        \hline
     
\cellcolor{white!20}   RO--OT in \cite{Efficient-OT-Naor}&\cellcolor{gray!50}   288 &\cellcolor{gray!50}    384\\

       \hline

\cellcolor{white!20}   Supersonic OT&\cellcolor{gray!20}    \textbf{0.75}&\cellcolor{gray!20}     1\\ 

    \hline

\end{tabular}
\end{center}
\end{table}


\begin{table}[!htbp]
\vspace{-3mm}
\caption{ The table compares the runtime of Supersonic OT with the following  general OT (G--OT) extensions: G--OT in  \cite{AsharovL0Z13} and G--OT in \cite{SchneiderZ13}. The runtime has been measured in milliseconds and is for $10^{\st 7}$ invocations of $1$-out-of-$2$ OT. The enhancement ratio refers to the performance improvement that Supersonic OT offers in comparison to each scheme.}  \label{fig::supersonic-vs-extension-OTs}
\begin{center}

\renewcommand{\arraystretch}{1.5}

\begin{tabular}{|c|c|c|c|c|c|c|c|c|c|c|c|} 
    \cline{1-4}

   Scheme&   Sec. Param. Size&   Runtime&   Enhancement Ratio\\

       \hline

\cellcolor{white!20}  G--OT in \cite{AsharovL0Z13}&  80-bit&\cellcolor{gray!50}    14,272&\cellcolor{gray!50}   2 \\
    
        \hline

\cellcolor{white!20}  G--OT in \cite{SchneiderZ13}&  80-bit&\cellcolor{gray!20}   20,717&\cellcolor{gray!20}   3 \\

       \hline

\cellcolor{white!50}  Supersonic OT&  128-bit& \cellcolor{gray!50}   \textbf{6,610}&\cellcolor{gray!50}    1\\ 

    \hline

\end{tabular}
\end{center}
\end{table}

Note that Supersonic OT maintains a consistent runtime across different security parameters, whether lower (e.g., 80-bit) or higher (e.g., 256-bit) than 128-bit. In contrast, the runtime of the schemes in \cite{AsharovL0Z13,Efficient-OT-Naor} would vary, fluctuating by at least a factor of 2.5 with changes in the security parameter.

\subsubsection{Supersonic Versus OT Extensions.}

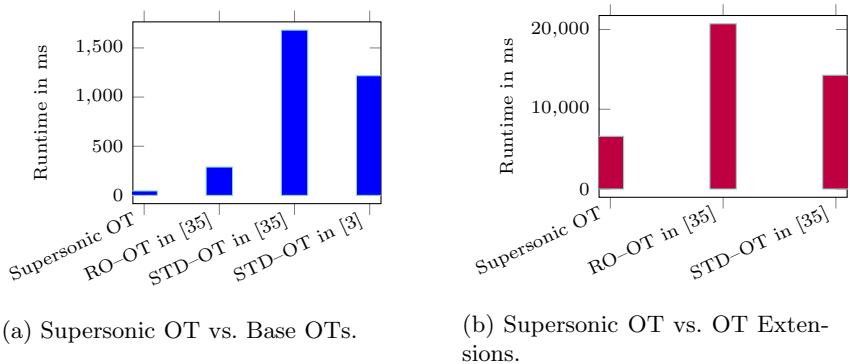
\begin{figure}[!htbp]
    \centering
    \begin{subfigure}[b]{0.4\textwidth}
        \centering
        \begin{tikzpicture}
            \begin{axis}[
                ybar,
                width=\linewidth,
                height=4cm,
                xlabel={},
                ylabel={\scriptsize Runtime in ms},
                ymin=0,
                xtick=data,
                xticklabels={
                    \scriptsize Supersonic OT,
                    \scriptsize RO--OT in \cite{Efficient-OT-Naor},
                    \scriptsize STD--OT in \cite{Efficient-OT-Naor},
                    \scriptsize STD--OT  in \cite{AsharovL0Z13}
                },
                xticklabel style={rotate=25, anchor=east},
                yticklabel style={font=\scriptsize},
                legend style={at={(0.5,-0.15)}, anchor=north, legend columns=-1},
                enlargelimits=0.05, 
                cycle list/Set3,
                ]
                \addplot+[fill=blue] coordinates {
                    (1, 45)
                    (2, 288)
                    (3, 1681)
                    (4,1217)
                };
            \end{axis}
        \end{tikzpicture}
        \caption{{\small{Supersonic OT vs. Base OTs.}}}\hspace{4mm}
        \label{supersonic-vs-base}
    \end{subfigure}%
    \hspace{12mm}
    \begin{subfigure}[b]{0.4\textwidth}
        \centering
        \begin{tikzpicture}
            \begin{axis}[
                ybar,
                width=\linewidth,
                height=4cm,
                xlabel={},
                ylabel={\scriptsize Runtime in ms},
                ymin=0,
                xtick=data,
                xticklabels={
                    \scriptsize Supersonic OT,
                    \scriptsize RO--OT in \cite{Efficient-OT-Naor},
                    \scriptsize STD--OT in \cite{Efficient-OT-Naor}
                },
                xticklabel style={rotate=25, anchor=east},
                yticklabel style={font=\scriptsize},
                legend style={at={(0.5,-0.15)}, anchor=north, legend columns=-1},
                enlargelimits=0.05, 
                cycle list/Set3,
                scaled y ticks=false,
                ]
                \addplot+[fill=purple] coordinates {
                    (1, 6610)
                    (2, 20717)
                    (3, 14272)
                };
            \end{axis}
        \end{tikzpicture}
        \caption{Supersonic OT vs. OT Extensions.}
        \label{supersonice-vs-extensions}
    \end{subfigure}
    \caption{Comparison of Runtimes for Different OTs.}
    \label{fig:improved_bar_chart_dynamic_scaling}
\end{figure}

We proceed to compare the runtime of Supersonic OT with the runtime of efficient OT extensions presented in \cite{AsharovL0Z13,SchneiderZ13}. Given that OT extensions are designed for scenarios involving frequent executions, we evaluate the runtime of these three OTs when invoked $10^{\st 7}$ times. 
The runtime data for OTs in \cite{AsharovL0Z13,SchneiderZ13} is extracted from the figures reported in \cite{AsharovL0Z13}, specifically from Tables 3 and 4 in \cite{AsharovL0Z13}, where the GMP library was utilized.\footnote{Table 4 in \cite{AsharovL0Z13} excludes the runtime of base OTs. Thus, the corresponding runtime of the base OT in Table 3 in \cite{AsharovL0Z13} must be added to each figure reported in Table 4. For instance, for G-OT (in the LAN setting) we would have $13.92\times 1000+ 352=14,272$.} Table \ref{fig::supersonic-vs-extension-OTs} and Figure \ref{supersonice-vs-extensions} present the outcome of this comparison.

Table \ref{fig::supersonic-vs-extension-OTs} shows that invoking  Supersonic OT $10^{\st 7}$ times takes approximately 6,610 ms with a 128-bit security parameter.



Supersonic OT outperforms the OT in \cite{AsharovL0Z13} by a factor of 2 and \cite{SchneiderZ13} by a factor of 3. Despite the higher 128-bit security parameter in Supersonic OT compared to the 80-bit parameter in the other two schemes, its runtime is still lower. We expect that increasing the security parameter in schemes in \cite{AsharovL0Z13,SchneiderZ13} would result in higher runtimes, given that the base OT's runtime increases accordingly.


\subsection{Features} 

For the base OTs and OT extensions in \cite{AsharovL0Z13,Efficient-OT-Naor,SchneiderZ13} to achieve unconditional security, as discussed in Section \ref{sec::uncon-sec-OT}, they typically require multiple replicas of the database, a noisy channel, or the involvement of a trusted initializer, all of which contribute to increased deployment costs. In contrast, Supersonic OT attains unconditional security without relying on database replications, noisy channels, or a fully trusted party. Although Supersonic OT involves an additional party, unlike base OTs or OT extensions that typically only involve the sender and receiver, it maintains its security even when this party is semi-honest.

\section{Conclusion and Future Work}

OT is a crucial privacy-preserving technology.  OTs have found extensive applications in designing secure Multi-Party Computation protocols \cite{Yao82b,AsharovL0Z13,HarnikIK07}, Federated Learning \cite{YangLCT19,RenYC22,XuLZXND22}, and accessing sensitive field elements of remote private databases while preserving privacy \cite{CamenischDN09,AielloIR01,libert2021adaptive}.

In this work, we presented Supersonic OT, an unconditionally secure very efficient $1$-out-of-$2$ OT that will maintain security in the quantum computing era. Supersonic OT refrains from using any public-key cryptography.  We have proved its security within the standard simulation-based paradigm. 

Supersonic OT enables the receiver to obtain a response of size $O(1)$ whose bit size can be as small as $128$ if the bit size of the secret messages maintained by the sender is at most 128. It relies on basic standard primitives. The simple design of Supersonic OT allows easy security analysis and implementation. 
We have studied the concrete performance of Supersonic OT. A single execution of Supersonic OT completes in approximately $0.35$ milliseconds, making it up to $2000$ times faster than the state-of-the-art base OT. Moreover, we studied the performance of Supersonic OT and the state-of-the-art OT extensions in the setting where they are invoked $10^{\st 7}$ times. Our analysis indicated that Supersonic OT takes approximately 6,610 ms, making it up to 3 times faster than current OT extensions.

As a future research direction, it would be interesting to see how the exceptional efficiency of Supersonic OT can improve the performance of those (generic MPC or Private Set Intersection) protocols that heavily rely on OT protocols.

\bibliographystyle{splncs04}
\bibliography{ref}

\end{document}